\documentclass[3p,twocolumn, times]{elsarticle}

\usepackage{ecrc}

\volume{00}

\firstpage{1}

\journalname{Nuclear Physics B Proceedings Supplement}

\runauth{}

\jid{nuphbp}

\jnltitlelogo{Nuclear Physics B Proceedings Supplement}

\usepackage{amssymb}

\usepackage[figuresright]{rotating}

\begin{document}

\begin{frontmatter}



\dochead{}

\title{The DAE$\delta$ALUS Experiment}


\author{Janet M. Conrad, for the DAE$\delta$ALUS Collaboration}

\address{Massachusetts Institute of Technology, Cambridge, MA 02139}

\begin{abstract}
  DAE$\delta$ALUS is an experiment to measure the $CP$-violation angle
  in the neutrino sector by producing multiple, intense beams of
  neutrinos from pion- and muon-decays-at-rest near an ultra-large
  water Cerenkov detector. In this talk, a design for the proposed
  Deep Underground Science and Engineering Laboratory in the U.S. was
  presented.  DAE$\delta$ALUS will be statistics-limited and have
  different systematic errors than long baseline $CP$-violation
  searches.  When the data from both searches are combined, the
  sensitivity exceeds proton-driver designs. In this proceeding, we
  briefly describe one of several alternative cyclotron designs under
  consideration for DAE$\delta$ALUS, as an example.

\end{abstract}


\end{frontmatter}



\section{Introduction to the Physics}

The physics community has placed the search for evidence for
$CP$-violation in the neutrino sector at the highest priority
\cite{APSNeutrinoStudy, NUSAG, P5}.  A primary reason is that
$CP$-violation in the light neutrino sector is considered a key piece
of evidence for the theory of leptogenesis \cite{BorisPDG}.  In this
theory, the light neutrinos are Majorana and have GUT-scale
partners.  The the matter-antimatter asymmetry of the universe is
explained through $CP$-violating decays of the heavy partners. It is
widely thought unlikely that $CP$-violation could appear in the heavy
sector without the occurrence in the light neutrinos \cite{BorisPDG}.
Thus, $CP$-violation in the neutrino sector is consider an important
``smoking gun.''

Beyond this, the search for $CP$-violation in the leptons is motivated
by the experimental observation of $CP$-violation in quarks.  In many
broad aspects, including flavor mixing, the leptons weak interactions
seem to mirror the quarks.  So it would, therefore, be very surprising
that $CP$ violation would be exactly zero in the lepton sector while
non-zero in quarks.  But the quark-lepton correspondence is like a
fun-house mirror -- the actual value of the parameters in the two
sectors are very different.  For example, mixing in the neutrinos is
very large compared to mixing in the quarks, but the neutrino masses
are very small compared to quarks.  It is important to ask if the $CP$ violating
parameter in neutrinos is similarly much larger than in quarks.
Understanding the patterns can give us bottoms-up clues to the
underlying theory of mass and flavor mixing in of particle physics.
The value of the leptonic $CP$-violation parameter is a key missing clue.

The search for a nonzero $CP$ violation parameter, $\delta$, requires a
neutrino oscillation appearance experiment.  At this point, the only
viable appearance experiment is muon flavor neutrinos oscillating to
electron flavor.  If we say that $\Delta_{ij}=\Delta
m_{ij}^{2}L/4E_{\nu}$ are the squared mass splittings and
$\theta_{ij}$ are the mixing angles parameterizing the oscillation,
then, neglecting matter effects, the oscillation probability is given
by \cite{Parke}:
\begin{eqnarray}
P  & =&\sin^{2}\theta_{23}\sin^{2}2\theta_{13}\sin^{2}%
\Delta_{13} \nonumber \\
& \mp&\sin\delta_{cp}\sin\theta_{13}\cos\theta_{13}\sin2\theta_{23}\sin2\theta _{12} \nonumber \\
&& ~~~~~~~~~~~~~~~\times \sin^{2}\Delta_{13}\sin\Delta_{12} \nonumber \\
& +&\cos\delta_{cp}\sin\theta_{13}\cos\theta_{13}\sin2\theta_{23}\sin2\theta 
_{12} \nonumber \\
&&  ~~~~~~~~~~~~~~~\times\sin\Delta_{13}\cos\Delta_{13}\sin\Delta_{12} \nonumber \\
& +&\cos^{2}\theta_{23}\sin^{2}2\theta_{12}\sin^{2}\Delta_{12},\label{equ:beam}
\end{eqnarray}
In this equation, $-(+)$ refers to neutrinos (antineutrinos).

With the exceptions of $\theta_{13}$ and $\delta$, the parameters in
this equation are known now, with improvements expected in the near
future \cite{schwetz,schwetz05,huber}.  Global fits to
$\sin^{2}(2\theta_{13})$ yield, at present, $0.06\pm 0.04$.  Reactor
experiments in the near future \cite{DayaBay, DoubleChooz} will push
the uncertainty down to 0.005.  The value of $\delta$ is entirely
unconstrained and measurement of this parameter is the goal of
DAE$\delta$ALUS.

Sensitivity to $\delta$ can arise in two ways.  One can take data with
neutrino and antineutrino beams and use the sign flip in
Eq.~\ref{equ:beam} to isolate $\delta$.  Or one can exploit the $L/E$
dependence of the interference terms
($\sin^{2}\Delta_{13}\sin\Delta_{12}$ and
$\sin\Delta_{13}\cos\Delta_{13}\sin\Delta_{12} \nonumber $).  This
second method allows sensitivity with either a neutrino or an antineutrino
beam.  Conventional searches use the former, while DAE$\delta$ALUS makes use
of the latter, and combined searches use both.

The above oscillation probability is valid in a vacuum or for a
short-baseline experiment, like DAE$\delta$ALUS.  Long baseline
experiments, such as those which use conventional neutrino beams, face
the complication of matter effects which depend on the unknown sign of
the mass hierarchy \cite{Parke}.

\section{Introduction to DAE$\delta$ALUS}

\begin{figure}[t]\begin{center}
{\includegraphics[width=3.in]{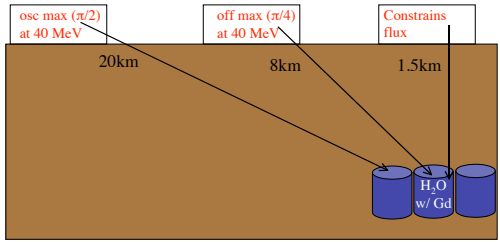}} 
\end{center}
\vspace{-0.25in}
\caption{Schematic of the layout of DAE$\delta$ALUS.  The powers at the respective sites, are, on average over the 10 year run, 1 MW, 2 MW and 5 MW
\label{layout} }
\end{figure}

\begin{figure}[t]\begin{center}
\vspace{0.45in}
{\includegraphics[width=3.in]{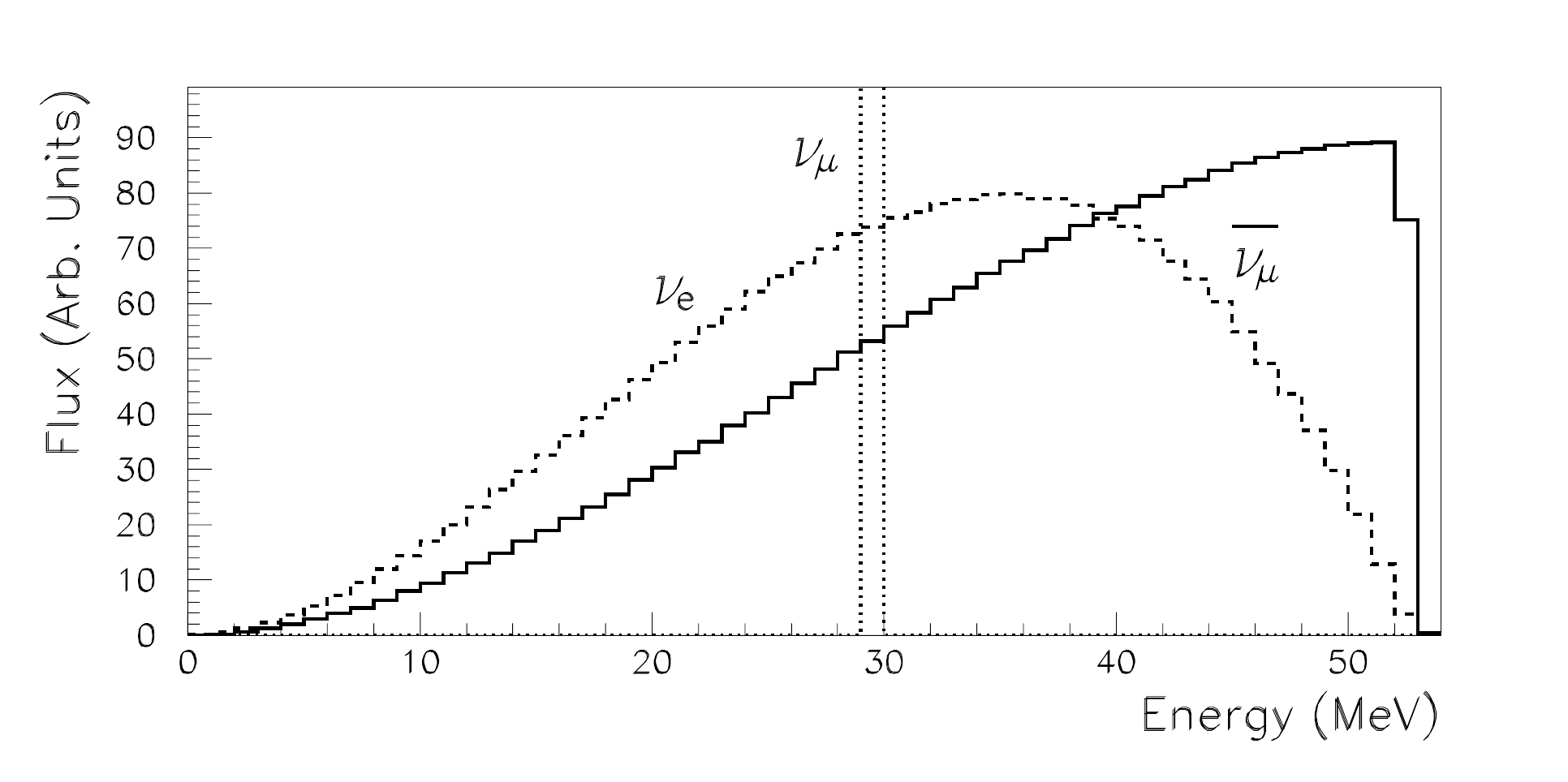}} 
\end{center}
\vspace{-0.25in}
\caption{Energy distribution for each flavor of neutrino in a decay-at-rest beam.
\label{flux} }
\end{figure}

The DAE$\delta$ALUS design is described in Refs.  \cite{EOI,
  firstpaper} and Fig.~\ref{layout}.  Cyclotrons are used to produce
pion and muon decay-at-rest neutrino beams. The search is for $\bar
\nu_\mu \rightarrow \bar \nu_e$ oscillations, exploiting the
length-dependence of the $CP$-violating interference terms in the
oscillation formula, Eq.~\ref{equ:beam}, to isolate $\delta$.  Three
cyclotron sources are used to map the oscillation as a function of
$L$. 

We propose to run this experiment at the new Deep Underground Science
and Engineering Laboratory (DUSEL) in South Dakota, although in
principle it can be installed near any ultra-large detector with free
protons.  The 1.5 km location is on the surface above the water
Cerenkov detector located at the 4850 level.  The 
other two sites are at 8 km and 20 km.  Given the low beam energy,
this preserves the $L/E$ necessary to be sensitive to the atmospheric
$\Delta m^2$.

This design relies on the fact that the weak decay chain produces beams with
identical energy dependence at each location.  Cyclotrons are planned
as a cost-effective, high-intensity method of producing protons to
create pions that lead to these decay-at-rest neutrino beams.  The 800
MeV protons impinge on a carbon target producing pions from the
$\Delta$ resonance.  These come to a stop in the target and
subsequently decay via the chain: $\pi^{+}\rightarrow\nu_{\mu}
\mu^{+}$ followed by $\mu^+ \rightarrow e^{+}\bar{\nu}_{\mu}\nu_{e}$.
The resulting flux, shown in Fig.~\ref{flux}, is isotropic with a well
known energy dependence for each of the three flavors.  Because almost
all $\pi^-$ capture before decay, the $\bar \nu_e$ fraction in the
beam is $\sim 4\times 10^{-4}$.  

The $CP$-violation search utilizes the $\bar \nu_\mu \rightarrow \bar
\nu_e$ channel.  The $\bar \nu_\mu$ flux is peaked towards the
endpoint of 52.8 MeV.  At the atmospheric $\delta m^2$, a 50 MeV beam
yields oscillation maximum at 20 km.  We will use three neutrino
sources located at 1.5 km (near), 8 km (mid) and 20 km (far).  The
cyclotrons at the three sites are run in alternating periods, so that
the $L$ for any given is event is known by the timing. The neutrinos
impinge on a single ultra-large detector with free-proton targets.
This allows us to use the inverse beta decay interactions (IBD), $\bar \nu_e
+ p \rightarrow n + e^+$, to identify $\bar \nu_e$ in the beam.  The
$\nu$-electron and $\nu_e$-oxygen scatters
are used to determine the relative normalization between sites.

To observe the $n$ capture in the IBD events with high efficiency, the
water Cerenkov detector would need to be Gd-doped.  Gd offers two
essential advantages over the competing process of hydrogen capture:
it reduces the capture time for the neutron generated in the IBD
interaction from 200 $\mu$s to about 30 $\mu$s , and the energy of the $\gamma$s
released from the capture interaction is higher - $\sim$8 MeV total 
(with about $\sim$5 MeV converted to observable Cerenkov light from Compton-scattered
electrons) compared
with a 2.2 MeV $\gamma$ for hydrogen.

\begin{figure}[t]\begin{center}
\vspace{-0.25in}
{\includegraphics[width=3.in]{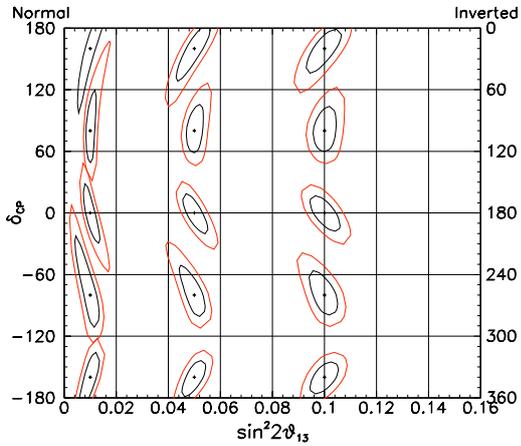}}
\vspace{-.85in}
\end{center}
\caption{Measurement capability of DAE$\delta$ALUS.   Inner region is 1$\sigma$, outer is 2$\sigma$.   There is an inherent ambiguity between the hierarchies,
with $\delta$ measured with a normal assumption being equivalent to $180^\circ - \delta$ in an inverted assumption.  
\label{deltasig} }
\end{figure}

The capability for measuring $\delta$ as a function of $\sin^2
2\theta_{13}$ is presented in Fig.~\ref{deltasig}.  DAE$\delta$ALUS
has an inherent ambiguity between the hierarchies because of its short
baseline, and this is indicated by the left and right $y$-axis scales
which correspond to normal and inverted hierarchies, respectively. For
making a discovery of $CP$-violation ($\delta$ not equal to 0$^\circ$
or 180$^\circ$) this ambiguity is not important.  The capability of
the experiment depends upon the tonnage of the H$_2$O detector.
Ref.~\cite{design} presents a study of DAE$\delta$ALUS capability as a
function of Gd-doped water Cerenkov detector mass.  Here, we present
results for 300 kt.

\section{Combining DAE$\delta$ALUS with a Conventional Beam}

\begin{figure}[t]\begin{center}
{\includegraphics[width=3.in]{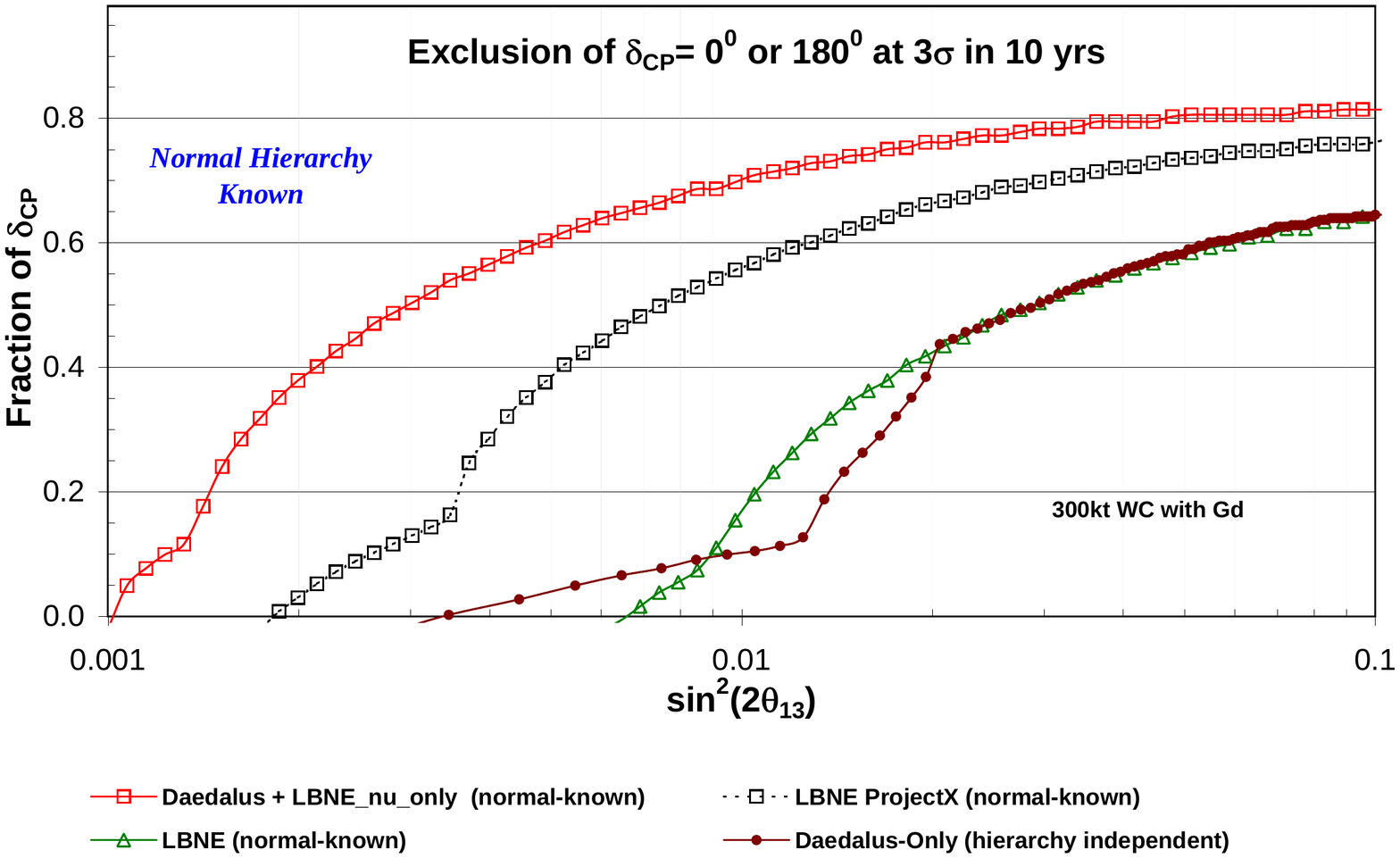}}
\vspace{-.5in}
\end{center}
\caption{Fraction of $\delta$-space where $\delta = 0^\circ$ or $180^\circ$
at 3$\sigma$ in 10 years, assuminga known normal hierarchy.    Red -- LBNE +
DAE$\delta$ALUS Combined, Black -- Project X, Green -- LBNE alone, Brown -- 
DAE$\delta$ALUS alone
\label{known} }
\begin{center}
{\includegraphics[width=3.in]{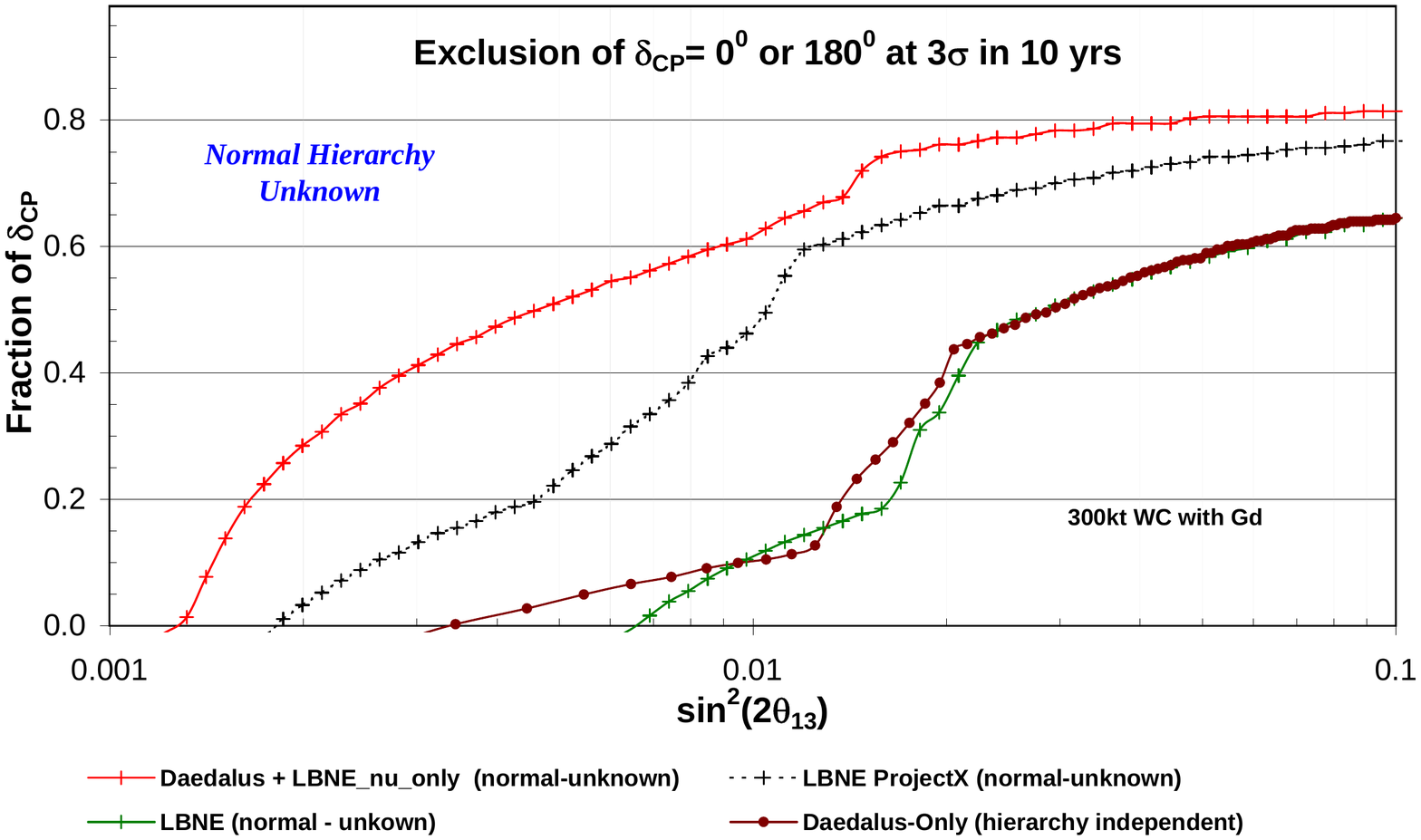}} 
\end{center}
\vspace{-0.5in}
\caption{An unknown mass hierarchy affects LBNE capability but  
does not affect DAE$\delta$ALUS directly.  This plot shows the
fraction of $\delta$-space where $\delta = 0^\circ$ or $180^\circ$
at 3$\sigma$ in 10 years, for hierarchy unknown to LBNE but is, 
in fact, normal.    Red -- LBNE +
DAE$\delta$ALUS Combined, Black -- Project X, Green -- LBNE alone, Brown -- 
DAE$\delta$ALUS alone
\label{unknown} }
\end{figure}

The DAE$\delta$ALUS data complements data from conventional-beam
searches for $CP$-violation, yielding high sensitivity when the data
sets are combined.  As an example, here we consider the LBNE neutrino
beam, which has a 1300 km baseline \cite{LBNE, Diwan} initiated at
Fermi National Accelerator Laboratory.  This beam is produced by 120 GeV
protons impinging on a target to produce pions and kaons.  These
mesons are magnetically focused, and subsequently decay to neutrinos
or antineutrinos, depending on the sign of the focusing field.  The
beam energy extends from about 300 MeV to above 10 GeV, and so is
sensitive to the same $L/E$ range as DAE$\delta$ALUS.

We consider four variations of beams impinging on the 300 kt water target,
with 10 year running-periods:
\begin{itemize}

\item \textbf{LBNE alone} -- which is 30$\times10^{20}$ protons on target (POT) in neutrino mode
followed by 30$\times10^{20}$ POT in antineutrino mode.  This is the standard
10-year run-plan for LBNE \cite{Mayly, Gina} prior to the startup of ``Project X.''

\item \textbf{DAE$\delta$ALUS alone} -- which is strictly antineutrino running, as described
above and  following the plan described in Ref. \cite{EOI}.

\item \textbf{Combined} -- which is the standard plan
  for DAE$\delta$ALUS antineutrino running combined with only-neutrino
  running of LBNE for the full 10 years.  This design builds on the
  strength of conventional beams, which is to produce pure and
  powerful neutrinos beams.  Conventional antineutrino beams can
  produce only about one-third the neutrino intensity and suffer from
  a high neutrino contamination \cite{LBNE}.  The DAE$\delta$ALUS and
  LBNE programs can take data simultaneously.

\item \textbf{Project X} -- which is a proposal for an upgrade to
  a ``proton driver'' which will yield ultra-high numbers of protons
  on target \cite{PX1, PX2}.  A standard expectation assumes the LBNE
  conventional beam with $10^{22}$ POT in 5 years in neutrino mode and
  $10^{22}$ POT for 5 years in antineutrino mode.

\end{itemize}

As an example of the relative capabilities, Fig.~\ref{known} shows the
fraction of $\delta$-space determined to be non-zero covered in the
various scenarios. LBNE alone (green) and DAE$\delta$ALUS alone
(brown) have roughly the same capability, by design.  What is
particularly impressive is that the combined capability (red) exceeds
that of Project X \cite{PX1, PX2} (black), as was also shown in
Ref.~\cite{Agar}, and reaches to very low values of $\sin^2
2\theta_{13}$.

LBNE is sensitive to the mass hierarchy.  Fig.~\ref{known} assumes a
known normal hierarchy, which seems plausible by the time these
experiments run.  However, if the hierarchy is unknown then LBNE
sensitivity is degraded, while DAE$\delta$ALUS sensitivity is not.
Using an underlying normal hierarchy as an example, Fig.~\ref{unknown}
shows the effect of an unknown hierarchy on the sensitivity for the
four scenarios.

\section{The Cyclotrons: An Example Design}

While superconducting linacs provide the most conservative accelerator
technology option, space and cost constraints suggest that it would be
best to develop high-power cyclotrons to meet our goals.  Several
desirable aspects inherent to cyclotrons attract us to this option.
First is compactness, minimizing costs for shielding and space, of
particular value for the near site where footprint will be an
important consideration.  Second is that the fixed-energy and
continuous beam character of cyclotrons are desirable features,
reducing peak-power loads on targets.

Three potential cyclotron designs under consideration for
DAE$\delta$ALUS were presented at the Cyclotrons 2010 conference
\cite{Luciano, Alonso}.  Here we focus on only one of the three
designs, for lack of space in these proceedings.  This is a design which is
particularly powerful and which is under development as an technology
for Accelerator Driven Systems (ADS) for Thorium Reactors.

\begin{figure}[t]\begin{center}
{\includegraphics[width=3.in]{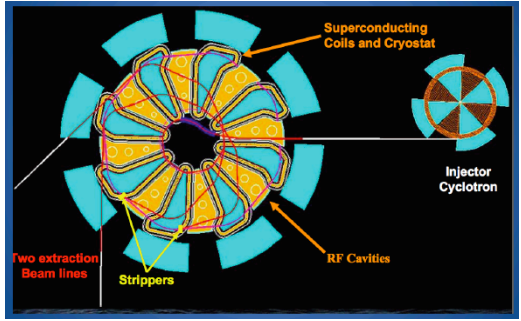}} 
\end{center}
\vspace{-0.25in}
\caption{Preliminary layout of a DAE$\delta$ALUS MMC.
\label{MMC} }
\end{figure}

The Multi-MegaWatt Cycoltron (MMC) design consists of an injector and
a booster cyclotron.  Specifics of the two components are given in
Table ~\ref{tab:MMC}. This machine accelerates H$_2^+$ ions which
provides advantages with respect to space charge effects and very
clean extraction through stripping foils.  The extraction goal is $>$
99.9\% efficiency, as has been achieved at PSI.  The design goal of
this machine is 12 mA of beam at 800 MeV, which can be compared to the
PSI experiment, which operates at 2.3 mA and 590 MeV.  This very high
instantaneous power is necessary for ADS operations.  However, for
DAE$\delta$ALUS application, the near accelerator will run run 67 ms
of 500 ms, the mid accelerator will run 133 ms of 500 ms and the far
accelerator runs 100 ms of 500 ms.  Thus the average power is similar
to PSI.  There are two extraction lines per accelerator.  Fig.~\ref{MMC}
shows the cyclotron, with the beam injected from the right,
accelerated and stripped with two extraction paths.

An important goal is a cost-effective design which, as much as
possible, uses commercially available equipment.  The ion source is
expected to be adapted for the ECR Visible Ion Source at Catania,
while the injector cyclotron is expected to be a modified commercial
model.  The booster cyclotron will be a custom design with economy in
its simplicity.  Dumps are also expected to be simple graphite
targets, as described in Ref.~\cite{EOI}.  Several aspects of the
overall design keep costs low compared to existing machines of similar
energy range, including: no need for a bunch structure, a single-energy
design, and no need to extract to secondary beams or accelerators.

\bigskip%
\begin{table}[tbp] \centering
{\footnotesize
\begin{tabular}
[c]{|c|c||c|c|}\hline
\multicolumn{4}{|c|}{Injector Cyclotron} \\ \hline
$E_{inj}$	& 35 keV/n	   & $E_{max}$	& 50 MeV/n \\ \hline
$R_{inj}$	& 49 mm	   & $R_{ext}$	& 1800 mm \\ \hline
$\langle B \rangle$ at Rinj & 1.09 T & $\langle B \rangle$ at $R_{ext}$ &	1.15 T \\ \hline
sectors	 & 4	           & Accel. Cavities &	4 \\ \hline
RF	& 25 MHz	   & Harmonic	& 3rd \\ \hline
V-inj	& $>$70 kV    	   & V-ext	& 250 keV \\ \hline
$\Delta E$/turn	 &1.8 MeV & $\Delta R$ at $R_{ext}$ &	16 mm \\ \hline \hline
\multicolumn{4}{|c|}{Extraction \& Magnetic Channels} \\ \hline
Deflector Gap & 12 mm & Electric Field & $<$50 kv/cm \\ \hline \hline
\multicolumn{4}{|c|}{Booster Cyclotron} \\ \hline
$R_{inj}$ &	1.65 m	  &     $R_{ext}$ &     4.95 m \\ \hline
$\langle B \rangle$ at $R_{inj}$ &  1.09 T & $\langle B \rangle$ at 
                                    $R_{ext}$ & 2.02 T \\ \hline
RF Cavities	 &  Single gap     &   N Cavities	& 8 \\ \hline
RF 	 &    59 MHz	   & Harmonic	 & 6th \\ \hline
V-peak   &	300-700 kV	   & $\Delta E$/turn	& 5.6 MeV  \\ \hline
$\Delta R$ at $R_{inj}$	& $>$20 mm & $\Delta R$ at $R_{ext}$ & 3 mm  \\ \hline
\hline
\end{tabular}}
\caption{Parameters of the injector and booster cyclotrons \label{tab:MMC}}%
\end{table}%

\section{Conclusion}

The DAE$\delta$ALUS experiment provides a new approach to the search
for CP violation in the light neutrino sector, using $\bar \nu_\mu
\rightarrow \bar \nu_e$ oscillations at short baselines.  The beam is
produced by high-power cyclotrons.  The signal is inverse beta decay
interactions in the 300 kt fiducial-volume Gd-doped water Cerenkov
neutrino detector proposed for the Deep Underground Science and
Engineering Laboratory.  This design could be employed at other
underground laboratories with ultra-large detectors.

DAE$\delta$ALUS provides a high-statistics, low-background complement to
conventional long-baseline neutrino experiments, like LBNE. 
Because of the complementary designs, when DAE$\delta$ALUS
antineutrino data are combined with LBNE neutrino data, the
sensitivity of the $CP$-violation search has sensitivity 
beyond the proposal for Project X. 

The experiment relies on beams produced by high-power cyclotrons which
are under development for commercial purposes.  Cyclotrons have the
advantage of being compact and low-cost compared to linacs. Three
designs are presently under consideration.  The example presented
here accelerates H$_2^+$ ions, which allows for very clean extraction
to multiple lines through electron stripping.  The consensus among
cyclotron physicists is that DAE$\delta$ALUS is a challenging and
interesting project, with high return on investment for both the
accelerator and neutrino communities.

\bibliographystyle{elsarticle-num}
\bibliography{Proceedings_Conrad_Janet_Saturday_morning}

\end{document}